\documentstyle[12pt]{article}
\title{\huge Monte Carlo Study of Order-Disorder Layering Transitions
in the Blume-Capel Model}
\author{\bf
 L. Bahmad, A. Benyoussef and H. Ez-Zahraouy$^{*}$
\\
 Laboratoire de Magn\'{e}tisme et de la Physique
 des Hautes Energies
\\
Universit\'{e} Mohammed V, Facult\'{e} des Sciences, Avenue Ibn Batouta,\\
Rabat  B.P. 1014, Morocco
}
\date{ }
\begin{document}
\maketitle

\begin{abstract}
\mbox{~~~  }

The order-disorder layering transitions, of the Blume-Capel model, are
studied using the Monte Carlo (MC) simulations, in the presence of a
variable crystal field. For a very low temperature, the results are in
good agreement with the ground state study. The first order transition line,
 found for low temperatures, is connected to the second order transition
line, seen for higher temperatures, by a tri-critical point, for each layer.
The reentrant phenomena, caused by a competition of thermal fluctuations and
an inductor
magnetic field  created by the deeper layers, is found for the first $k_0$
layers from the surface, where $k_0$ is exactly the number of layering
transitions allowed by the ground state study. The layer magnetizations $m_k$,
the magnetic susceptibilities $\chi_{m,k}$ and the quadrupolar magnetic
susceptibilities $\chi_{q,k}$, for each layer $k$, are also investigated.

\end{abstract}

----------------------------------- 
\mbox{~}(*) Corresponding author: ezahamid@fsr.ac.ma \\

\newpage

\section{Introduction}
\mbox{  }
Since the theory of surface critical phenomena started developing, much attention
has been devoted to the study of the Blume Capel (BC) model over semi-infinite
lattices, with modified surface couplings. Benyoussef {et al.} [1] have
 determined the phase diagram in the mean field approximation, reporting
 four possible topologies at fixed bulk/surface coupling ratios. An analoguous
 analysis have also been done using a real space renormalization group
 transformation [2]. Some other works referring to particular regions of the
 phase space are, for example, those using: the mean field approximation [3], the
 effective field approximation [4] and the low temperature expansion [5].
 These works show that it is possible to have a phase with ordered
 surface and disordered bulk, which is separated from the completely ordered
 phase by the so-called {\it extraordinary} transition and from the
completely disordered phase by the {\it surface} transition.
In the absence of this phase, the transition between the completely ordered and the
completely disordered phase is called {\it ordinary}.
These three kinds of phase transitions have a meeting point named {\it special} and
which is generally a multi-critical point.
The discussion presented in Ref. [6] shows that the strong interest in these
models arises partly from the unusually rich phase transition behavior they
display as their interaction parameters are varied, and partly from their many
possible applications.
The bilinear interaction considered in most of these cases is ferromagnetic.
The spin-$1$ Ising systems are used to describe both the order-disorder
transition and the crystallisation of the binary allow, and it was solved in
the mean field approach [7].
The decomposition of a line of tri-critical points into a line of critical
end points and one of double critical points is is one of the most interesting
and elusive features of the mean field phase diagram for the anti-ferromagnetic
spin-$1$ Blume-Capel model in an external magnetic field [8].
The transfer-matrix and Monte Carlo finite-size-scaling methods [9],
are also applied to study this model but such decomposition does not occur
in the two dimensional case.
The finite cluster approximation has been applied by Benyoussef {\it et al.}
[10] in order to study the spin-$1$ Ising model with a random crystal field. \\
On the other hand, the transverse field or crystal field effects
of spin-$1$ Ising model has been studied by several authors [11-14]. \\
The experimental measurements of layer-by-layer ordering phenomena have been
established on free-standing liquid crystals films such as ${\it nm}OBC$
(n-alkyl-4'-n-alkyloxybiphenyl-4-carboxylate) [15,16] and $54COOBC$ (
n-pentyl-4'-n-pentanoyloxy-biphenyl-4-carboxylate) [17] for several
molecular layers. More recently, Lin {\it el al.} [18] have used the
three-level Potts model to show the existence of layer-by-layer ordering
of ultra thin liquid crystal films of free-standing $54COOBC$ films, by
adjusting the interlayer and intra layer couplings between nearest-
neighboring molecules.  \\
The reentrant first-order layering transitions have been observed
experimentally in multilayer argon films on graphite by Youn and Hess [19].
Using the mean field method we have shown recently, see Ref. [20],
the existence of order-disorder layering transitions in the Blume-Capel Ising
films under the effect of a variable crystal field according to the law
$\Delta_k=\Delta_s/k^\alpha$ ($\Delta_s$ being the surface crystal field,
$k$ the layer number counted from the surface and $\alpha$ a positive
constant).
Our aim in this paper is to study the model we introduced in Ref. [20],
using Monte Carlo simulations in order to examine the layer-by-layer
order-disorder transitions, the existence of the reentrant phenomena for each
layer, in one hand, and to investigate the magnetic layer susceptibilities and
the corresponding critical exponents. \\
This paper is organized as follows. Section $2$ describes the model and the method.
In section $3$ we present results and discussions. \\

\section{Model and method}
The system we are studying here is formed with $N$ coupled ferromagnetic
square layers in the presence of a crystal field. The Hamiltonian governing
this system is given by
\begin{equation}
{\cal H}=-\sum_{<i,j>}J_{ij}S_{i}S_{j}+\sum_{i}\Delta_{i}(S_{i})^2
\end{equation}
where, $ S_{l}(l=i,j)=-1,0,+1$ are the spin variables. The interactions
between different spins are assumed to be constant so that $J_{ij}=J$.
The crystal field acting on a site $i$ is so that $\Delta_{i}=\Delta_{k}$
for all spins of the layer $k$ so that:
$\Delta_{1}>\Delta_{2}>...>\Delta_{N}$. \\
The model we are studying, in this paper, corresponds to a crystal field
distributed according to the law:
\begin{equation}
\Delta_{k}=\Delta_{s}/k^{\alpha}
\end{equation}
where $\Delta_s=\Delta_{1}$ is the crystal field acting on the surface
(first layer $k=1$) and $\alpha$ a positive constant. \\

The quantities computed, using the Monte Carlo simulations for each layer
$'k'$ containing $N_x$ spins in $x-$direction and $N_y$ spins in the
$y-$direction, are: \\
$\bullet$ The layer average magnetizations $m_k=<M_k>$, where the site average
magnetization $M_k$ of the layer $k$ is given by
\begin{equation}
M_k=(\sum_{i\epsilon k}{S_i})/(N_x  N_y)
\end{equation}
$\bullet$ The layer magnetic susceptibilities defined as
\begin{equation}
	\chi_{m,k}=\beta N_x  N_y  <(M_k-m_k)^2>
\end{equation}
$\bullet$ The layer quadrupolar magnetic susceptibilities expressed as
\begin{equation}
\chi_{q,k}=\beta  N_x  N_y  <(q_k-<q_k>)^2>
\end{equation}
where
\begin{equation}
q_k=(\sum_{i\epsilon k}{S_i{^2}})/(N_x  N_y).
\end{equation}
$\bullet$ The layer critical exponents $\gamma_{m,k}$ related to
the corresponding layer magnetic susceptibilities $\chi_{m,k}$, for a
fixed layer 'k, giving by
\begin{equation}
\gamma_{m,k}=\frac{\partial Log(\chi_{m,k})^{-1}}
{\partial Log \mid T_{c,k}/J-T/J \mid}
\end{equation}
where $T/J$ stands for the absolute temperature and $T_{c,k}/J$ is the
critical temperature of the layer 'k'. \\
In the above equations $\beta=1/(k_B T)$ with $k_B$ being the Boltzmann
constant. \\

The notation $D^kO^{N-k}$ will be used to denote that the first $k$ layers
from the surface are disordered while the remaining ${N-k}$ layers are ordered.
 In particular, $O^N$ corresponds to an ordered film whereas $D^N$ denotes
 a totally disordered film. The surface crystal field $\Delta_{k}$, applied
 on each layer $k$ is distributed according to the law given by Eq. $(2)$.

\section{Results and discussion}
 \mbox{  } Monte Carlo simulations have been made on a system with $N$
 layers and $N_x=N_y$ spins for the $x$ and $y-$directions of each layer.
Runs of $500 000 / N$ Monte Carlo steps (MCS) were performed with the discard
of the first $50 000 / N$ Monte Carlo steps. \\
We have found that by changing the system sizes from $N=10$ to $N=15,20$
layers (and $N_x=N_y=64$ to $N_x=N_y=128$), the relevant measured quantities
did not change appreciably. For this reason we give in all the following,
numerical results for a film formed with $N=10$ layers and $N_x=N_y=64$ spins for
the directions $x$ and $y$. \\
The ground state phase diagram of this model was established in Ref. [20].
It is shown that, for very small values of the surface crystal field
$\Delta_s$, the system orders in the phase $O^N$. When increasing $\Delta_s$,
the surface (first layer $k=1$) disorders and the phase $DO^{N-1}$ occurs at
$\Delta_s/J=3(1)^\alpha$. Increasing $\Delta_s$ more and more, the second
layer $k=2$ becomes disordered at $\Delta_s/J=3(2)^\alpha$, and so on.
The transition from the phase $D^kO^{N-k}$ to the phase $D^{k+1}O^{N-(k+1)}$
 is seen at $\Delta_s/J=3(k+1)^\alpha$ provided that $k+1 \le N$.
For higher values of the surface crystal field the system is totally
disordered and the phase $D^N$ occurs. \\
In particular, we showed the existence of a critical order layer $k_0$
corresponding to the transition $D^{k_0}O^{N-k_0} \leftrightarrow D^N$, at
a reduced surface crystal field given by:
\begin{equation}
\Delta_s/J=(3(N-k_0)-1)/(\sum_{k=k_0+1}^{N}(1/k^\alpha)).
\end{equation}
$k_0$ is exactly the number of layering transitions existing at $T/J=0$.
It is found that $k_0$ depends both on the parameter $\alpha$ and the film
thickness $N$. The special case: $\alpha=0$ is a situation with a constant
crystal field applied on each layer and there is only a single transition
$O^N \leftrightarrow D^N$, occurring at $\Delta_s/J=3-1/N$ for $T/J=0$. \\

In order to outline the existence of the order-disorder layering transitions
we plot in Fig. 1 the temperature-crystal field phase diagram
 for $\alpha=1.0$.
 For non null but very low temperatures,
the surface crystal field values found in this phase diagram are exactly
those predicted by the ground state study established in Ref. [20], namely
: $\Delta_s/J=3(k)^\alpha$ for the first $k_0$ layers. Indeed, the first layer transition is found at
$\Delta_s/J=3.0$; the second transition at $\Delta_s/J=6.0$; the third
transition at $\Delta_s/J=9.0$ and so on.
Monte Carlo simulations show that the last layers $k=8,9$ and $10$
transit simultaneously at the numerical value of the surface crystal field:
$\Delta_s/J \approx 23.84$.
This is in good agreement with the predicted value $(\approx 23.80)$
from the above Eq. $(8)$ established on the basis of an analytical
study. \\
The reentrant phenomena, shown in Fig. $1$ for the first $k_0$ layers, is
caused by the competition between thermal fluctuations and an inductor
magnetic field created by the deeper layers. Indeed, when these thermal
fluctuations become sufficiently important, the magnetization of some spins,
of deeper layers, becomes non null (+1 or -1). This leads to the appearance
of an inductor magnetic field.
This magnetic field is responsible of the ordered
phase seen for the layer $k$.
This argument can also explain the absence of the
reentrant phenomena for the last layers, once the magnetization of the
remaining $N-k$ layers, is not sufficient to create an inductor magnetic
field. It is worth to note that the reentrant phenomena is always present
for the layers $k$, ($k \leq k_0$), and the corresponding tri-critical points
$C_i$ are located at a constant temperature. \\
In absence of the surface crystal field, $\Delta_s/J \rightarrow 0$, each
layer of the film disorders at a fixed temperature $T_c/J = 3.15$.
In order to compare this critical temperature value obtained for $N=10$ layers,
with the three-dimensional spin-1 critical temperature, we performed Monte Carlo
simulations on films with a fixed surface $(N_x,N_y)=(128,128)$ and increasing
thicknesses: $N=16;32;64$ and $128$. It is found that the corresponding critical
temperature increases very slowly with the number of layers once $N \ge 32$.
For $N=128$, the critical temperature we computed did not exceed the value
$T_c/J = 3.19(5)$. This value is smaller than that one computed by the mean field
approch for a three-dimensional cubic lattice spi-$1$ Ising model, which
is exactly $T_c/J = 4.0$.\\
On the other hand, as it is shown in Fig. 1, the first $k_0$ tri-critical
temperature temperature values, established by MC simulations are approximtely:
$T_{ci}/J \approx 0.50$. These temperature values are smaller than those
established in Ref. [20] when using the mean field method: $T_{ci}/J \approx 0.78$.
A detailed study is done for a very low temperature, as it is shown in
Figure 2 for $T/J=0.2$. Indeed, Fig. $2a$ shows that the individual layering
transition temperatures undergo a first order transition at the crystal
field values predicted by the analytical study, namely: $\Delta_s/J=3.0$ for
the first layer $k=1$, $\Delta_s/J=6.0$ for the second layer $k=2$,
$\Delta_s/J=9.0$ for the third layer $k=3$, and so on. The last layers $k=8$,
$k=9$ and $k=10$ transit simultaneously at $\Delta_s/J \approx2 3.84$.
On the other hand the layer quadrupolar magnetic susceptibilities $\chi_{q,k}$
defined by Eq. $(5)$, for each layer $k$, present a strong peak at the
surface crystal field values
corresponding to each layer transition; as it is illustrated by Fig. $2b$.
For a higher temperature, Fig. $3$ corresponding to $T/J=1.0$, the layering
transitions are second order type. Indeed, this is shown in Fig. $3a$ for
the individual layering temperature behaviors, as a function of the surface
crystal field. The corresponding magnetic susceptibilities $\chi_{m,k}$,
defined by Eq. $(4)$ for each layer $k$, present a strong peak at the
corresponding surface crystal field. It is worth to note that Figs.
$2b$ and $3b$ are plotted for reduced numerical values of the layer magnetic
and the layer quadrupolar magnetic susceptibilities: $\chi_{q,k}$ and
$\chi_{m,k}$ respectively.  \\
To show the increasing temperature effect, for a fixed surface crystal field,
on the order-disorder transitions; we illustrate in Fig. $4$ the
corresponding layering transitions for $\Delta_s/J=2.0$ for each layer. It is
found that these transitions are second order type and are
located at $T/J \approx 2.95$, see Fig. $4a$ for the layer magnetization behavior and Fig. $4b$ for the layer magnetic susceptibilities $\chi_{m,k}$.
On the other hand the reentrant phenomenon, seen for the first $k_0$ layers
is well illustrated in Fig. $5$ for the layer $k=7$ at a fixed surface
crystal field value $\Delta/ J=22.0$. Indeed, the two transitions:
from disorder to order and from order to disorder of the layer $k=7$ are
well seen in Fig. $5$ for the layer magnetization $m_7$. The corresponding
layer magnetic susceptibility $\chi_{m,7}$ exhibits two strong peaks at
these transitions.
In order to complete this study, we have investigated the layer critical
exponents $\gamma_{m,k}$ for $N=10$ layers and several system sizes $N_x \times N_y =16 \times 16$, $N_x \times N_y =32 \times 32$, $N_x \times N_y =64 \times 64$, and $N_x \times N_y =128 \times 128$; see Fig. 6. It is found that
$\gamma_{m,k}$ decreases, for a fixed order 'k', with the system size
$N_x \times N_y$ and stabilizes at certain value; and due to the free
boundary conditions the critical exponents of the layers $k=1$ and $k=N$
are found to be greater than those of the internal layers $2 \le k \le N-1$.

\section{Conclusion}
The order-disorder layering transitions of the Blume-Capel Ising model have
been studied under the effect of a variable crystal field
according to the law Eq. (2) and using Monte Carlo simulations.
The reentrant phenomena, caused by a competition of thermal fluctuations and
an inductor magnetic field  created by the deeper layers, is found for the
first $k_0$ layers counted from the surface; where $k_0$ is exactly the number
 of layering transitions allowed by the ground state study.
We established the temperature-crystal field phase diagrams and found that
the last $N-k_0$ layer tri-critical points are located at higher temperature
values for a fixed exponent $\alpha$ and film thickness $N$.
For very low temperatures, the results are in
good agreement with those predicted by the ground state study.
The first order and the second order transition lines are connected
by a tri-critical point, for each layer.
On the other hand, the layer magnetizations $m_k$,
the magnetic susceptibilities $\chi_{m,k}$, the quadrupolar magnetic
susceptibilities $\chi_{q,k}$, and the critical exponents $\gamma_{m,k}$
for each layer $k$, have been computed.

\newpage
\noindent{\bf References}
\begin{enumerate}

\item[{[1]}] A. Benyoussef, N. Boccara and M. Saber, J. Phys. C {\bf 19}, 1983 (1986).

\item[{[2]}] A. Benyoussef, N. Boccara and M. Bouziani, Phys. Rev. B {\bf 34}, 7775 (1986).

\item[{[3]}] X. P. Jiang and M.R. Giri, J. Phys. C {\bf 21}, 995 (1988).

\item[{[4]}] I. Tamura, J. Phys. Soc. Jpn {\bf 51}, 3607 (1982).

\item[{[5]}] C. Buzano and Pelizzola, Physica A {\bf 195}, 197 (1993).

\item[{[6]}] J.B. Collins, P.A. Rikvold and E. T. Gawlinski, Phys. Rev. B {\bf 38}, 6741 (1988).

\item[{[7]}] Y. Saito, J. Chem. Phys. {\bf 74}, 713 (1981).

\item[{[8]}] Y. L. Wang and K. Rauchwarger, Phys.Lett. A {\bf 59}, 73 (1976).

\item[{[9]}] J.D. Kimel, P. A. Rikvold and Y. L. Wang, Phys. Rev. B {\bf 45}, 7237 (1992).

\item[{[10]}] A. Benyoussef and H. Ez-Zahraouy, J. Phys.: Cond. Matt.  {\bf 6}, 3411 (1994).

\item[{[11]}] R. B. Stinchcombe, J. Phys. C {\bf 6}, 2459 (1973).

\item[{[12]}] J. L. Zhong, J. Liangli and C. Z. Yang, Phys. Stat. Sol. (b) {\bf 160}, 329 (1990).

\item[{[13]}] U. V. Ulyanoc, O. B. Zalavskii, Phys. Rep. {\bf 216}, 179 (1992)

\item[{[14]}] A. Benyoussef, H. Ez-Zahraouy and M. Saber, Phys. A {\bf 198}, 593 (1993)

\item[{[15]}] R. Geer, T. Stoebe, C. C. Huang, R. Pindak, J. W. Goodby, M. Cheng, J. T. Ho and S. W. Hui, {\it Nature} {\bf 355}, 152 (1992).

\item[{[16]}] T. Stoebe, R. Geer, C. C. Huang and J. W. Goodby, Phys. Rev. Lett. {\bf 69}, 2090 (1992).

\item[{[17]}] A. J. Jin, M. Veum, T. Stoebe, C. F. Chou, J. T. Ho, S. W. Hui, V. Surendranath and C. C. Huang, Phys. Rev. Lett. {\bf 74}, 4863 (1995);
Phys. Rev. E {\bf 53}, 3639 (1996).

\item[{[18]}] D. L. Lin, J. T. Ou, Long-Pei Shi, X. R. Wang and A. J. Jin, Europhys. Lett. {\bf 50}, 615 (2000).

\item[{[19]}] H. S. Youn and G.B. Hess, Phys. Rev. Lett. {\bf 64}, 918 (1990).

\item[{[20]}] L. Bahmad, A. Benyoussef and H. Ez-Zahraouy, J. Magn. Magn. Mat. {\bf 251}, 115 (2002).


\end{enumerate}

\noindent{\bf Figure Captions}\\

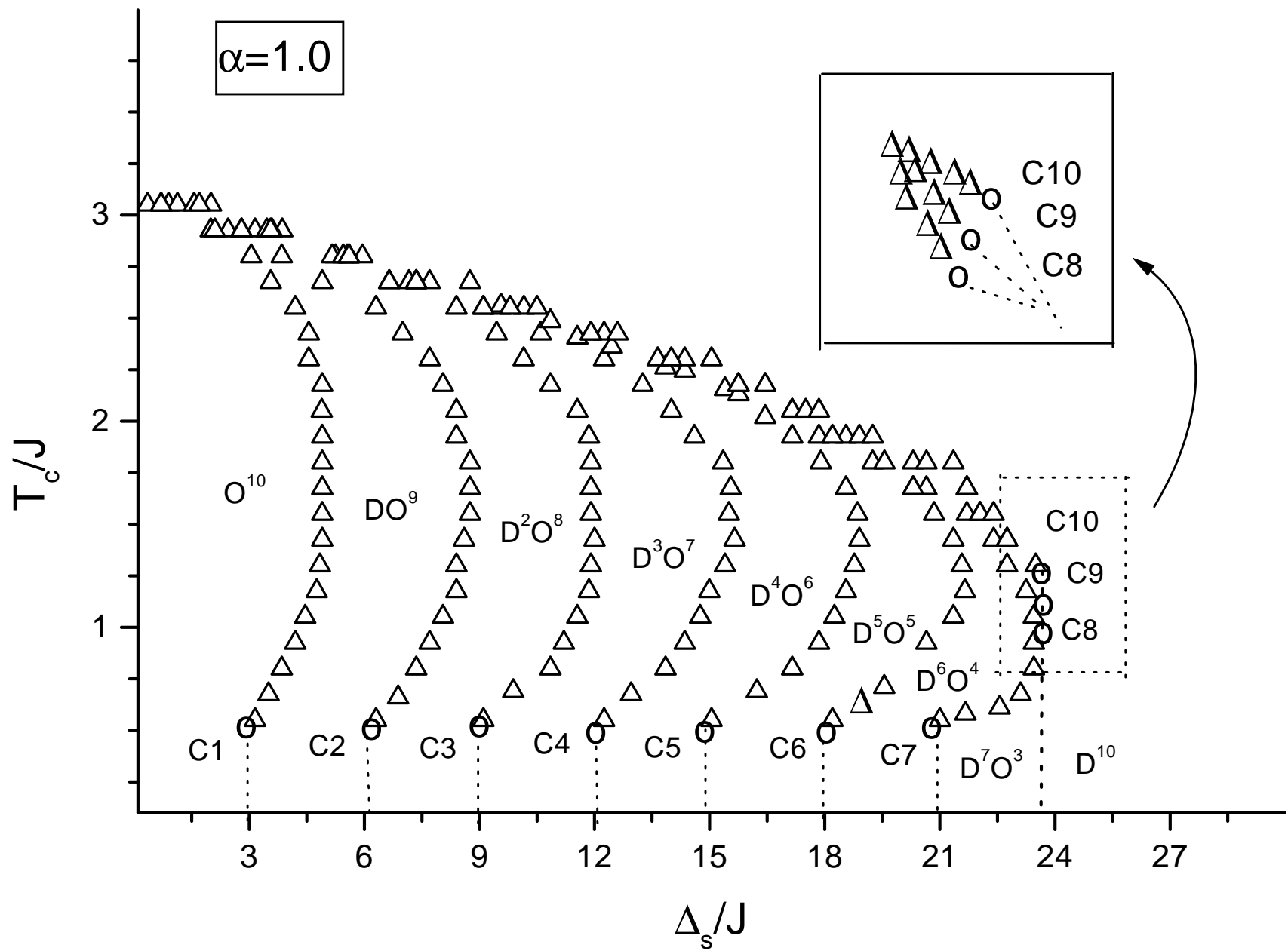
\noindent{\bf Figure 1.}: The critical temperature behavior as a function
of the surface crystal field $\Delta_s /J$ for $\alpha=1.0$ and a film
thickness $N=10$ layers. For each layer $'k'$ $(k=1,2,...,N)$, the first-order
transition line (vertical dashed line) is connected to the second-order
transition line (up-triangular points) by a tri-critical point $C_k$
(open circle). The notations $D^{k}O^{N-k}$ are defined in the body text.\\

\noindent{\bf Figure 2.}: The behavior of the reduced layer magnetizations
$m_k$ (Fig. a), and the layer quadrupolar magnetic susceptibilities
$\chi_{q,k}$ (Fig. b), showing a first-order transition, as a function of
the surface crystal field $\Delta_s /J$, for a very low temperature:
$T /J=0.2$.
The number accompanying each curve denotes the layer order from the surface
to deeper layers. \\

\noindent{\bf Figure 3.}: For a higher temperature $T /J=1.0$, the dependency
 of the layer magnetizations $m_k$ (Fig. a), and the reduced layer
magnetic susceptibilities $\chi_{m,k}$ (Fig. b), undergoes a second-order
transition, as a function of the surface crystal field $\Delta_s /J$. \\

\noindent{\bf Figure 4.}: For a fixed surface crystal field value
 $\Delta_s /J=2.0$, the layer magnetizations $m_k$ (Fig. a) and the reduced
 layer magnetic susceptibilities $\chi_{m,k}$ (Fig. b) show a second-order
transition as a function of increasing temperature. For each layer $'k'$
$(k=1,2,...,N)$, the transition is located at $T /J=2.95$. \\

\noindent{\bf Figure 5.}: Thermal behavior of the layer magnetization $m_7$
(line with solid circles) and the reduced layer magnetic susceptibility
$\chi_{m,7}$ (line with open circles) for a deeper layer $k=7$.
The reentrant phenomena as well as the second order transition are both
outlined by the layer magnetization transition from $D^{7}O^{3}$ to
$D^{6}O^{4}$ and from $D^{6}O^{4}$ to $D^{10}$ corresponding to two peaks
of the reduced layer magnetic susceptibility $\chi_{m,7}$ for
$\Delta_s/J=22.0$. \\

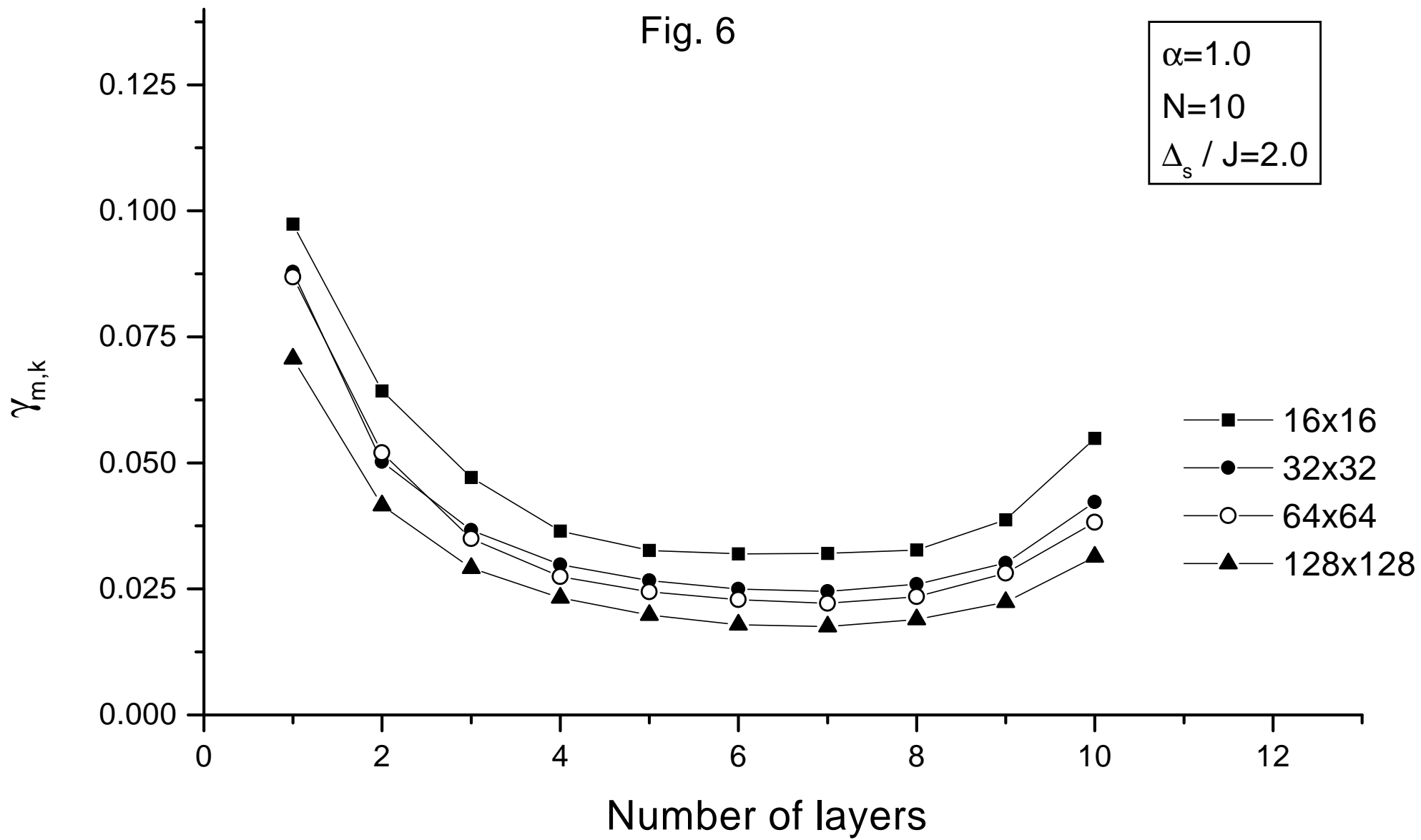
\noindent{\bf Figure 6.}: Layer critical exponents $\gamma_{m,k}$ for $\Delta_s/J=2.0$
as a function of the position 'k' for a film thickness $N=10$ layers, and several sizes :
$N_x \times N_y =16 \times 16$ (fill squares);
$N_x \times N_y =32 \times 32$ (fill circles);
$N_x \times N_y =64 \times 64$ (open circles);
and $N_x \times N_y =128 \times 128$ (filled up-triangulars).

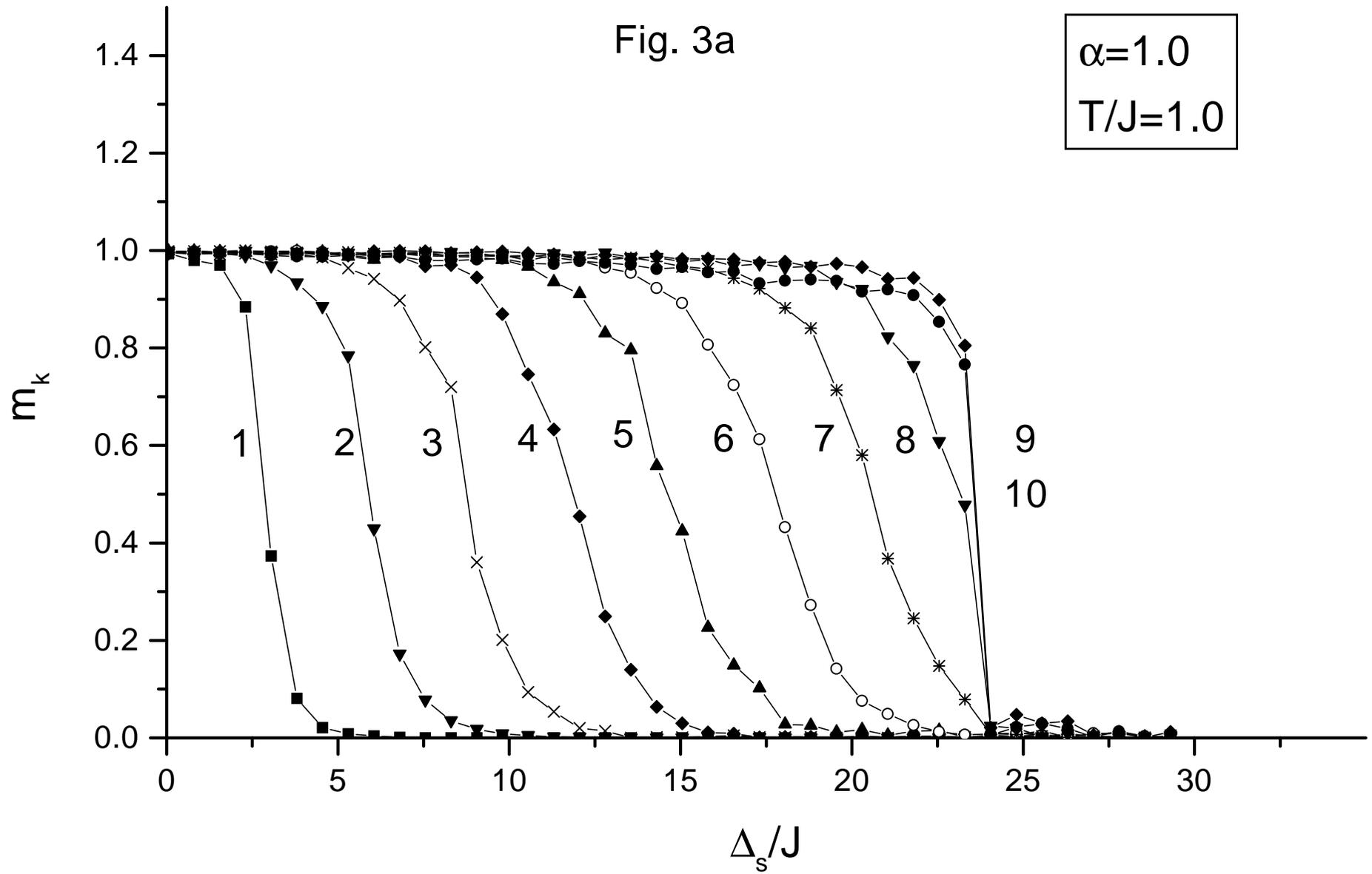
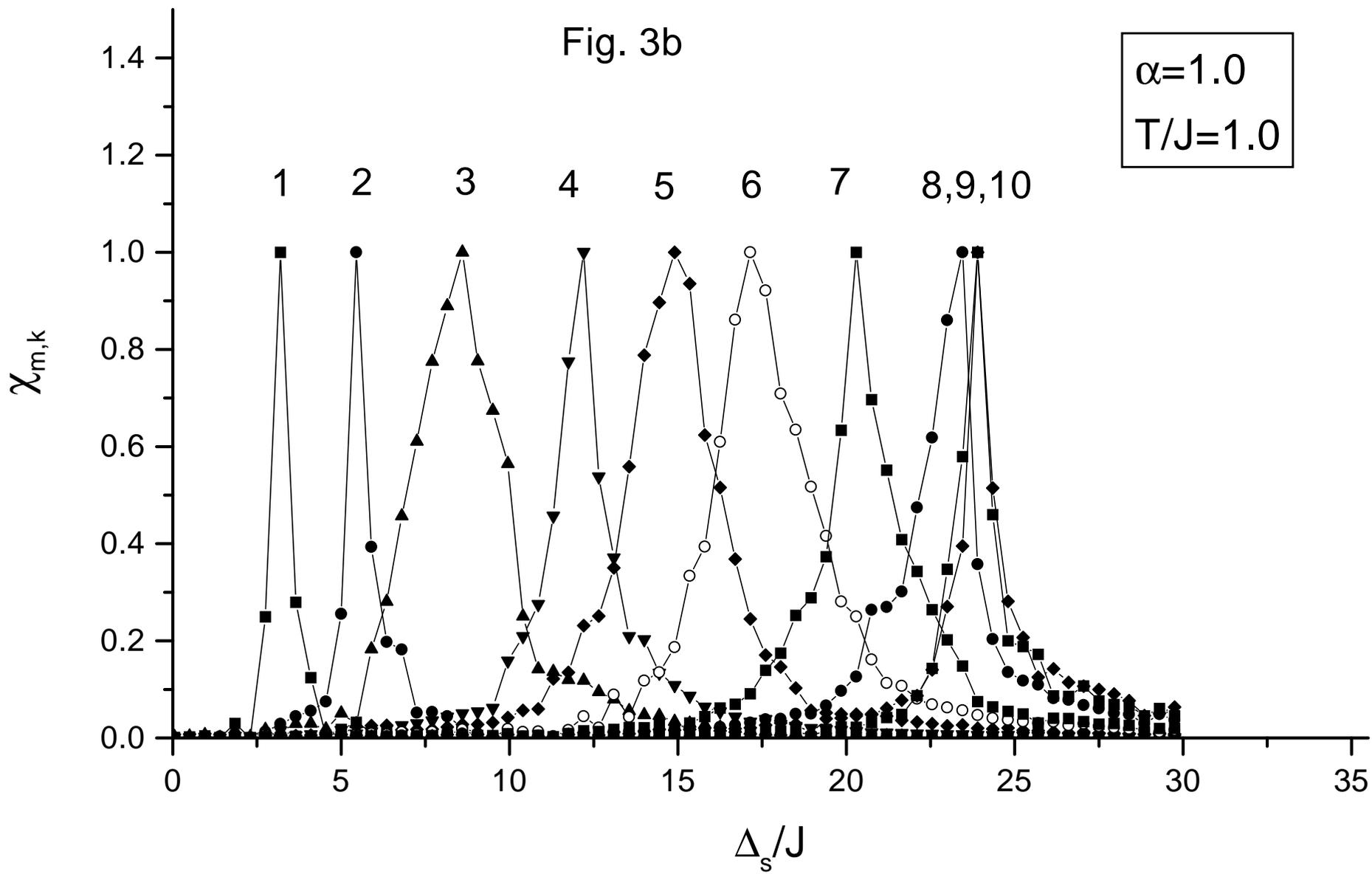
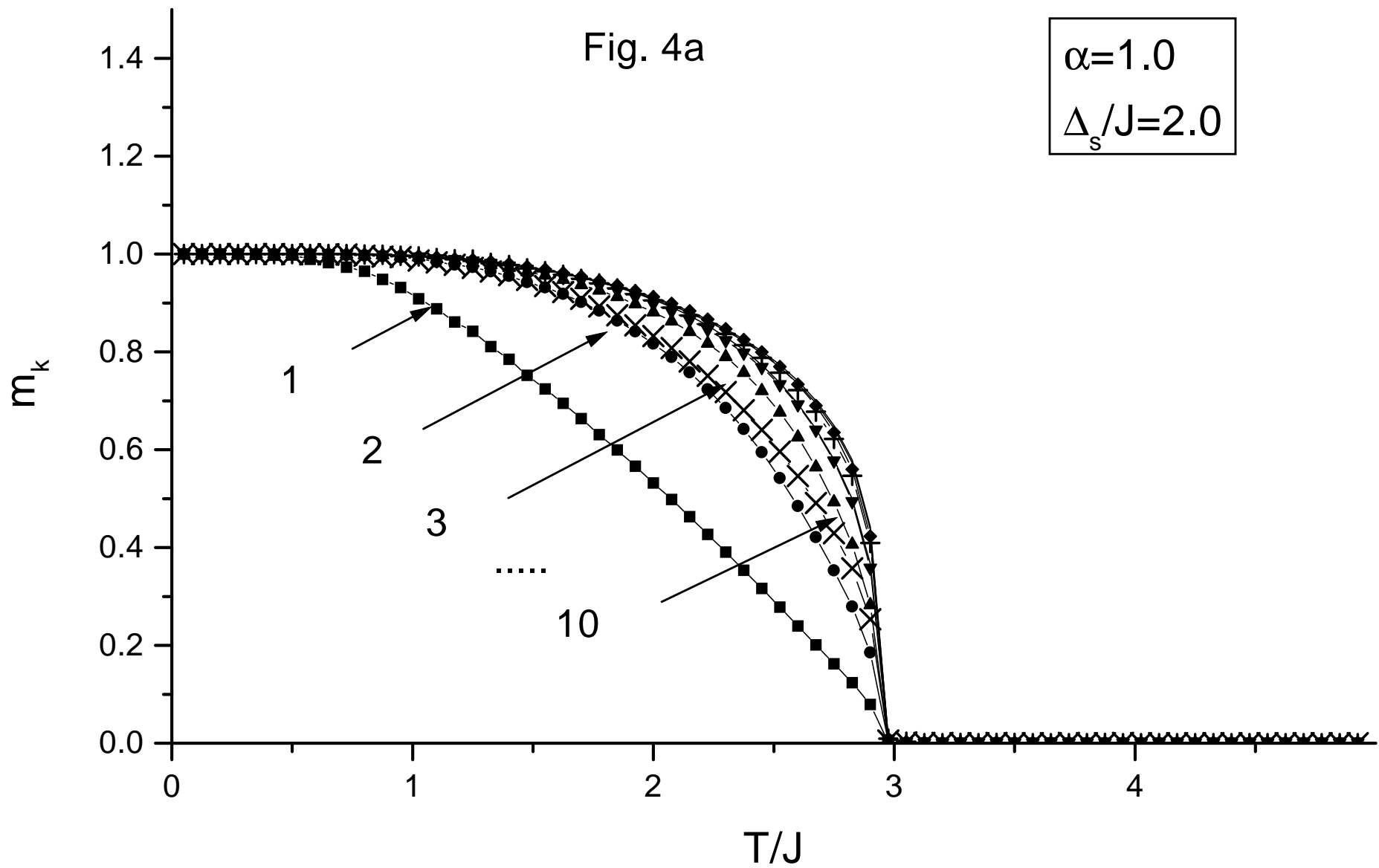
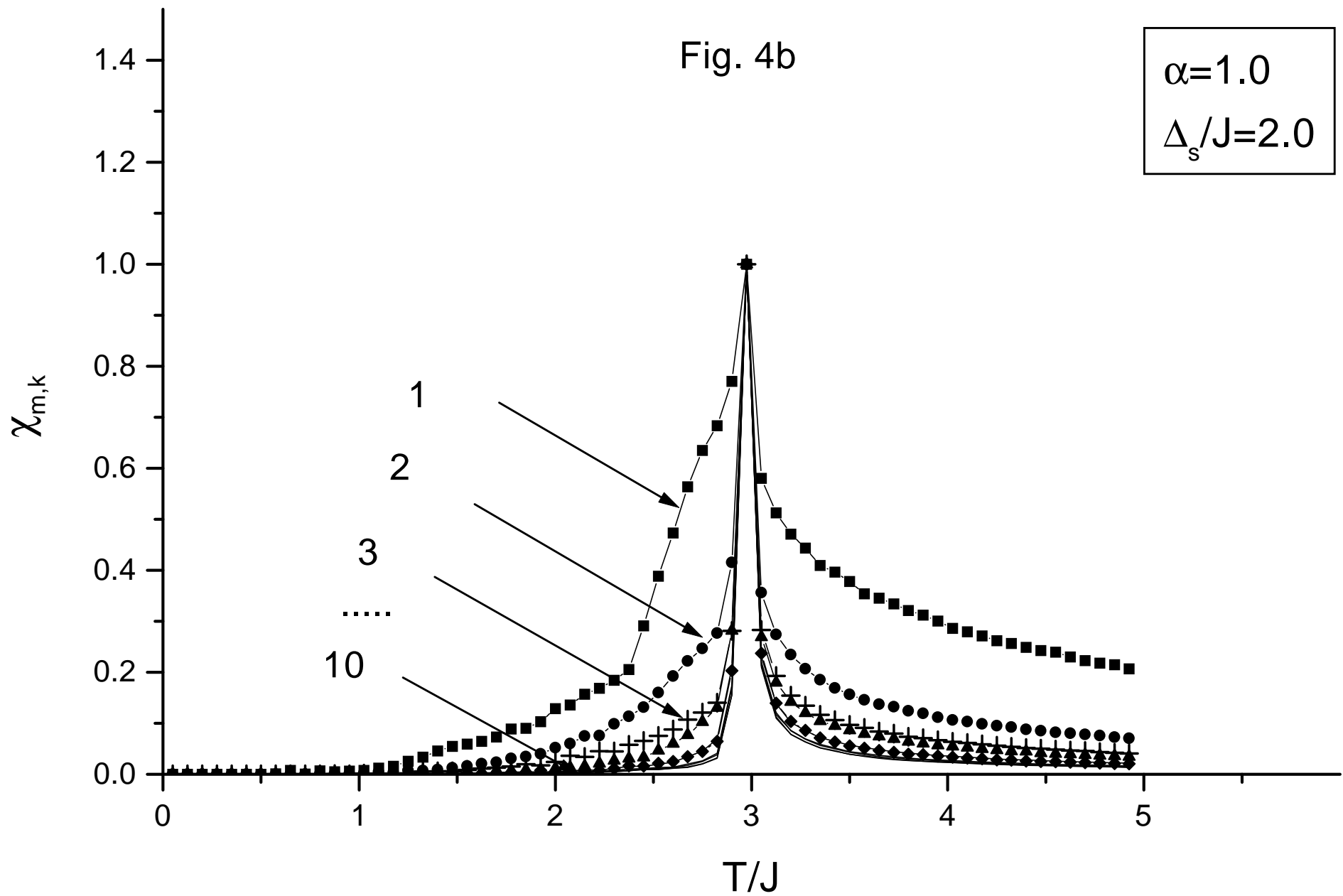
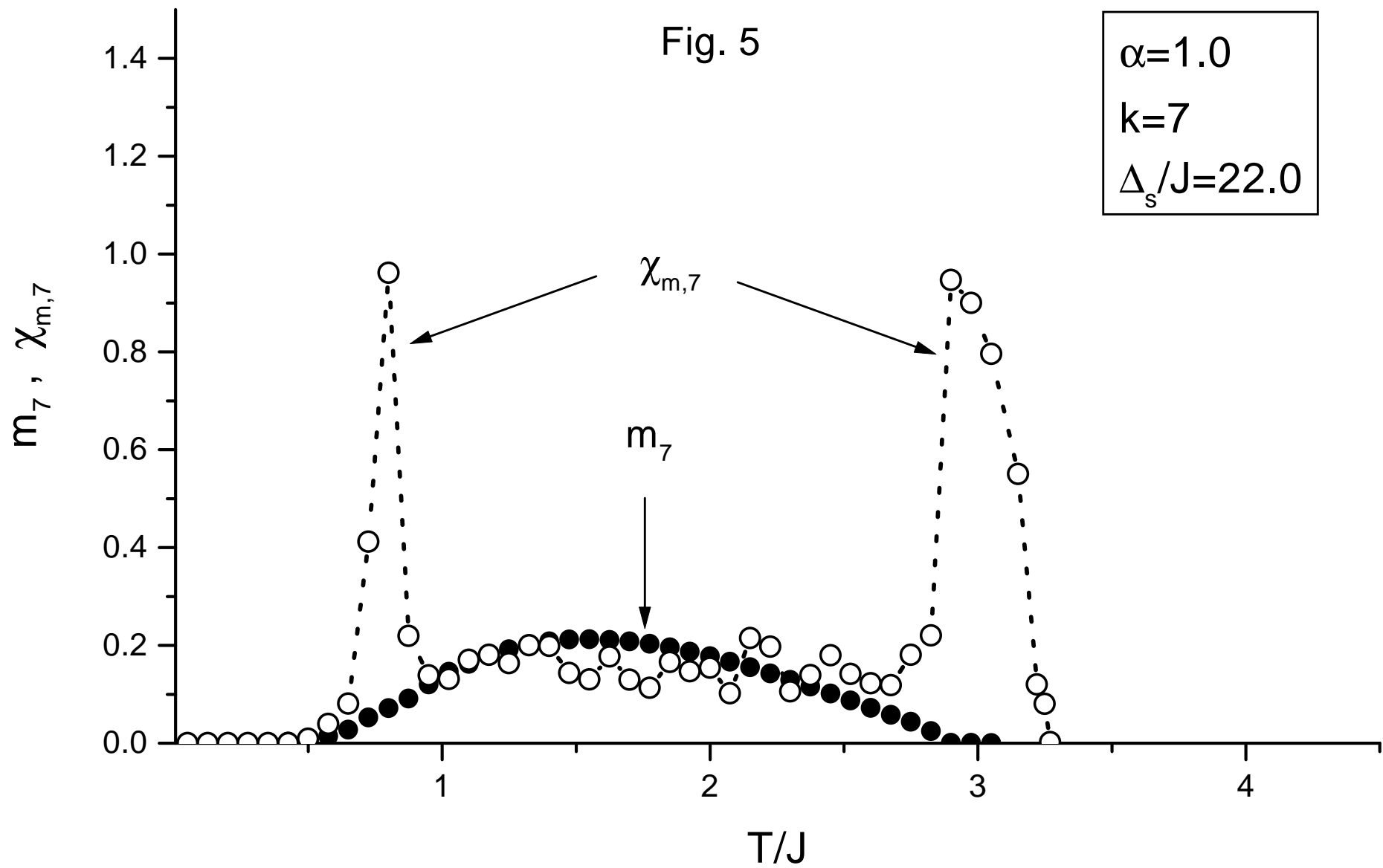
\end{document}